\documentclass[%
 aip,
 rsi,
amsmath,amssymb,
reprint,
floatfix
]{revtex4-1}

\usepackage{graphicx}
\usepackage{dcolumn}
\usepackage{bm}

\usepackage[utf8]{inputenc}
\usepackage{mathptmx}
\usepackage{multirow}
\usepackage{braket}
\newcommand{\ketbra}[2]{| #1\rangle\! \langle #2 |}

\usepackage[english]{babel}
\usepackage{xcolor}
\colorlet{RED}{red}
\colorlet{BLUE}{blue}
\usepackage{dcolumn}
\usepackage{bm}
\usepackage[version=4]{mhchem} 
\usepackage{acronym}
\usepackage{adjustbox}
\usepackage{float}
\usepackage{tikz}
\usepackage{microtype} 
\usepackage{algorithm}
\usepackage{algpseudocode}

\usetikzlibrary{calc,shapes.geometric,decorations.pathmorphing,patterns}

\definecolor{background-color}{gray}{0.98}
\usepackage[margin=2.3cm,bmargin=1cm,footnotesep=1cm]{geometry}

\begin{document}

\title{
Quantum algorithms for generator coordinate methods
}

\author{Muqing Zheng}
\affiliation{%
  Lehigh University, Bethlehem, Pennsylvania, 18015, USA
}
\affiliation{%
  Pacific Northwest National Laboratory, Richland, Washington, 99354, USA
}

\author{Bo Peng}
\affiliation{%
  Pacific Northwest National Laboratory, Richland, Washington, 99354, USA
}

\author{Nathan Wiebe}
\affiliation{%
  University of Toronto, Toronto, Ontario, M5G 1Z8, Canada
} 
\affiliation{%
  Pacific Northwest National Laboratory, Richland, Washington, 99354, USA
}

\author{Ang Li}
\affiliation{%
  Pacific Northwest National Laboratory, Richland, Washington, 99354, USA
}

\author{Xiu Yang}
\affiliation{%
  Lehigh University, Bethlehem, Pennsylvania, 18015, USA
}

\author{Karol Kowalski}
\email{karol.kowalski@pnnl.gov}
\affiliation{%
  Pacific Northwest National Laboratory, Richland, Washington, 99354, USA
 }

\begin{abstract}
This paper discusses quantum algorithms for the generator coordinate method (GCM) that can be used to benchmark molecular systems. The GCM formalism defined by exponential operators with exponents defined through generators of the Fermionic $U(N)$  Lie algebra (Thouless theorem) offers a possibility of probing large sub-spaces using low-depth quantum circuits. In the present studies, we illustrate the performance of the quantum algorithm for constructing a discretized form of the Hill-Wheeler equation for ground and excited state energies. 
We also generalize the standard GCM formulation to multi-product extension that when collective paths are properly probed, can systematically introduce higher rank effects and provide elementary mechanisms for symmetry purification when generator states break the spatial or spin symmetries. The GCM quantum algorithms also can be viewed as an alternative to existing variational quantum eigensolvers, where multi-step classical optimization algorithms are replaced by a single-step procedure for solving the Hill-Wheeler eigenvalue problem.
\end{abstract}

\maketitle

\section{Introduction}
\label{section1}

The rapid development of quantum technologies and quantum algorithms addresses long-standing computational challenges of many-body physics and quantum chemistry. While the primary target is to overcome exponential growth in complexity associated with approaching the exact limit in the simulations, the possibility of identifying physically meaningful solutions to problems of interest is equally important. Although Quantum Phase Estimation algorithms 
\cite{luis1996optimum,
cleve1998quantum,berry2007efficient,childs2010relationship,wecker2015progress,
haner2016high,poulin2017fast}
have been designed in a way that adequately addresses both of these issues, their applicability is currently limited by the necessity of using complex quantum circuits with corresponding depths that preclude its practical applications on existing quantum computing platforms, dominated by  Noisy Intermediate-Scale Quantum (NISQ) devices.
Instead, hybrid algorithms such as various Variational Quantum Eigensolver (VQE) 
\cite{peruzzo2014variational,mcclean2016theory,romero2018strategies,PhysRevA.95.020501,Kandala2017,kandala2019,PhysRevX.8.011021,huggins2020non,ryabinkin2018qubit,cao2019quantum,ryabinkin2020iterative,izmaylov2019unitary,lang2020unitary,grimsley2019adaptive,grimsley2019trotterized,cerezo2021variational,mcardle2020quantum,bharti2022noisy,anand2022quantum}
are currently being intensively tested on NISQ quantum computers to characterize the properties of correlated quantum systems.
In this effort, in many aspects, the VQE formulations, for example, based on the unitary coupled-cluster (UCC)\cite{unitary1,unitary2,hoffmann1988unitary,kutzelnigg1991error,evangelista2019exact} representation of the wave function, mirror the standard conventional formulations of CC theory, where a large number of excitations are included in the cluster operator and simultaneously optimized in the iterative process. This algorithm leads to another set of challenges associated with the potential problems with the convergence of iterative process (commonly referred to as the barren minimum problem) and representation of UCC Ansatz on quantum registers, which may result in long Trotter-like products of exponential operators defined by multi-qubit gates. Although several strategies mitigating these problems have recently been proposed, utilizing VQE-UCC formulations and extending these methods beyond specific system-size limits may be challenging and require reformulation of the quantum many-problem into recently introduced quantum flow equations,\cite{kowalski2021dimensionality} where large subspaces of Hilbert space can be sampled through the constant-depth small-dimensionality coupled eigenproblems.

Instead, in this paper, we explore the applicability of the Generator Coordinate Method (GCM) 
\cite{hill1953nuclear,rodriguez2002correlations,bender2003self,ring2004nuclear,yao2010configuration,egido2016state,hizawa2021generator}
as an alternative to popular VQE formulations. The main difference with the VQE method is that the GCM method avoids highly-nonlinear parametrization of the wave function and provides an efficient mean for direct extension of the probed subspaces. Additionally, it offers an efficient utilization of Ansatzes represented by low-depth quantum circuits. 

The GCM method was one of the first attempts to combine two distinct aspects of many-body theories: independent-particle models and theories describing collective phenomena, where the approximate eigenstates $|\Psi_{\rm GCM}\rangle$ of the Hamiltonian $H$ are expressed using a family of $N$-body wave functions $|\Phi({\bf q})\rangle$,
\begin{equation}
|\Psi_{\rm GCM}\rangle = \int d{\bf q} |\Phi({\bf q})\rangle f({\bf q}) \;,
\label{intro1}
\end{equation}
where ${\bf q}$ is a set of collective variables that describe correlation effects in many-body systems, and usually, the corresponding $|\Phi({\bf q})\rangle$ is represented a complicated linear combination of Slater determinants. The scalar $f({\bf q})$ is referred to as the weight function. The advantage of the GCM approach is obtaining ground states and classes of excited states described by the chosen set of generator coordinates. In general, in the analogy to the coherent state representation,\cite{klauder1985coherent,zhang1990coherent} the family $|\Phi({\bf q})\rangle$ forms an over-complete basis.
Upon substituting (\ref{intro1}) into Sch\"{o}dinger equation one gets the so-called Hill-Wheeler integral equations for unknown
$f({\bf q})$ coefficients
\begin{equation}
\int d{\bf q}’\lbrack {\bf H}({\bf q},{\bf q}’) – E {\bf S}({\bf q},{\bf q}’)\rbrack 
f({\bf q}’) = 0 \;,
\label{intro2}
\end{equation}
where the integral kernels ${\bf H}$ and ${\bf S}$ are defined as
\begin{eqnarray}
{\bf H}({\bf q},{\bf q}’) &=& \langle\Phi({\bf q})|H|\Phi({\bf q}')\rangle \;, \label{intro3} \\
{\bf S}({\bf q},{\bf q}’) &=& \langle\Phi({\bf q})|\Phi({\bf q}')\rangle \;. \label{intro4} 
\end{eqnarray}
In typical applications, the Hill-Wheeler equation is usually solved numerically by discretization, which transforms integral equation (\ref{intro3}) into an algebraic eigenvalue problem.

There are two categories of GCM formulations in applications to many-body quantum systems.
The first category's purpose is to restore broken symmetries of the $|\Phi({\bf q})\rangle$ states. For example, Bardeen-Cooper-Schrieffer states are not eigenstates of the particle number operator $N$. To project the Bardeen-Cooper-Schrieffer states onto the Hilbert space with the desired number of particles $N_0$, one uses the projection operator expressed in terms of integral over gauge angle $\phi_{N_0}$,
\begin{equation}
P_{N_0}=\frac{1}{2\pi} \int_{0}^{2\pi} d\phi_{N_0} e^{i\phi_{N_0} (N-N_0)} \;.
\label{intro5}
\end{equation}
In this case, the properties of $f_k({\bf q})$ coefficients are determined by the properties of the symmetry (projection) operators (e.g., particle number projection operator). In the second category of GCM formulations, the unknown weight function is optimized to capture correlation effects encoded in generator coordinates.
However, designing an adequate grid or path to probe the Hill-Wheeler equation is a rather empirical procedure that requires  much intuition and prior knowledge of the sought-after many-body system.

The appealing feature of the GCM method, especially from the point of possible quantum computing applications, is the possibility of combining low-depth representations of $|\Phi({\bf q})\rangle$ functions with simple, one-step optimization conditions for weight function. In this approach, the role of quantum computing is to map a discrete number of states
$\lbrace |\Phi({\bf q}_p)\rangle\rbrace_{p=1}^M$ to quantum register and to evaluate matrix elements for Hamiltonian and overlap matrices (Eqs. (\ref{intro3}) and (\ref{intro4})). In contrast, solving a generalized eigenvalue problem in a discrete basis representation takes place only once on a classical computer, avoiding multiple instances of quantum-classical communication as in the VQE formalism. In the discussed formalism, we follow an “algebraic” GCM formulation discussed by Fukutome in Ref.\onlinecite{fukutome1981group}, extend the standard GCM to multi-product exponential formulas, and provide an algorithm for sampling coordinate space in a way that provides selective approaching classes of excited Slater determinants.
In this context, the multi-product GCM formulation alleviates some problems associated with the usage of high-fidelity of Trotter-type expansions.

We demonstrate the performance of the quantum GCM algorithm on the example of the H4 benchmark system in various configurations. We
show that by a judicious choice of the $|\Phi({\bf q})\rangle$, one can recover high-level of accuracy both in weakly and strongly correlated regimes using $|\Phi({\bf q})\rangle$s that use the manifold of single excitations (or $U(N)$ Lie algebra generators). The obtained level of accuracy is similar to the one obtained with the advanced VQE formulations. 


\section{Theory}

The GCM was initially introduced to describe collective effects in nuclei.\cite{hill1953nuclear,goeke1980generator,ring2004nuclear,hizawa2021generator}
Fukutome in his seminal paper \cite{fukutome1981group} considered the GCM from the Lie-algebra theoretical standpoint. We use that language throughout this paper. 
Let us assume that the Fermion system is described by annihilation and creation operators $a_p$ and $a_q^{\dagger}$ that satisfy the following set of anticommutation relations: 
\begin{equation}
[a_p,a_q]_+=[a_p^{\dagger},a_q^{\dagger}]_+=0 \;,\;\;
[a_p,a_q^{\dagger}]_+ = \delta_{pq}
\label{antic}
\end{equation}
As discussed in Ref.\onlinecite{fukutome1981group} 
there are several Fermion algebras including 
$U(N)$, $SO(2N)$, and $SO(2N+1)$ Lie algebras, and Clifford algebras that can be used to characterized approximate many-body wave functions (here, $N$ stands for the number of single particle states). 
These algebras can be defined by the following set of operators 
\begin{equation}
E^p_q=a_p^{\dagger}a_q-\frac{1}{2}\delta_{pq} \;,\;
E_{pq}=a_p a_q \;,\; E^{pq}=a_p^{\dagger}a_q^{\dagger}\;.
\label{aux1}
\end{equation}
For example, 
\begin{itemize}
    \item $U(N)$: $\lbrace E^p_q \rbrace$ \;,
    \item $SO(2N)$: $\lbrace E^p_q, E_{pq},E^{pq} \rbrace$ \;,
    \item $SO(2N+1)$: $\lbrace a_p, a_q^{\dagger}, E^p_q, E_{pq}, E^{pq} \rbrace$ \;.
\end{itemize}
The $U(N)$ algebra is the only algebra where particle number operator commutes with all operators belonging to $U(N)$ algebra.  

Let us focus attention on the $U(N)$ algebra. 
By $\Gamma({\bf Z})$ we designate an anti-Hermitian operator defined as 
\begin{equation}
\Gamma({\bf Z})=\sum_{p,q} z_{pq} E^p_q =\sum_{p,q} z_{pq} a_p^{\dagger}  a_q
\;\;,\;\; z_{qp}^{\star}=-z_{pq}
\end{equation}
where ${\bf Z}$ can be viewed as an anti-Hermitian matrix $[z_{pq}]$ where $z_{qp}^{\star}=-z_{pq}$
or set of indexed parameter $\lbrace z_{pq} \rbrace$
satisfying $z_{qp}^{\star}=-z_{pq}$.
Canonical transformation $U({\bf Z})$ generated by $\Gamma({\bf Z})$ takes the form 
\begin{equation}
U({\bf Z})=e^{\Gamma({\bf Z})} \;.
\label{expz}
\end{equation}
Standard canonical Thouless transformation of the Hartree-Fock (HF) determinant $|\Phi\rangle$ can be obtained by acting with $U({\bf Z})$ onto $|\Phi\rangle$, i.e.,
\begin{equation}
|\Phi({\bf Z})\rangle = U({\bf Z})|\Phi\rangle =
e^{\Gamma({\bf Z})}|\Phi\rangle 
\label{unz}
\end{equation}
The standard unitary CC model with singles (UCCS) is a special case of (\ref{unz}) where 
$z_{ij}=z_{ab}=0$, where $i,j,\ldots$ and $a,b,\ldots$
correspond to spin-orbital indices occupied and unoccupied in the Slater determinant $|\Phi\rangle$
(where ${\bf Z}$ can be identified with the set of elements $\lbrace z_{ia} \rbrace$).
In comparison to $|\Phi\rangle$, 
$|\Phi({\bf Z})\rangle$ can contain certain classes of correlation effects (that are ${\bf Z}$ dependent; it can contain elements of symmetry breaking and effects responsible for the appearance of HF instabilities).


The parametrized states (\ref{unz}) can be viewed as a non-orthogonal 
(in general overcomplete) basis in the Hilbert space.
The generator coordinate method utilizes this fact by
representing the wave function in the form 
\begin{equation}
|\Psi_{\rm GCM}\rangle = \int d{\bf Z}  |\Phi({\bf Z})\rangle f({\bf Z}) \;,
\label{gcmexp}
\end{equation}
where collective variables ${\bf q}$ of Eq.(\ref{intro1}) are now identified with ${\bf Z}$ matrix, i.e., 
\begin{equation}
{\bf q} \rightarrow {\bf Z} \;.
\label{ident1}
\end{equation}

A typical way for solving Hill-Wheeler equations (\ref{intro2}) is through discretization of the ${\bf Z}$ domain. Let us 
introduce a set of the ${\bf Z}$-points - 
$Q=\lbrace {\bf Z}_i \rbrace_{i=1}^M$
then equations (\ref{intro2}) takes the form of non-orthogonal eigenvalue problem
\begin{equation}
{\bf H} {\bf f}=E {\bf S} {\bf f},
\label{eigendisc}
\end{equation}
where ${\bf H}$ and ${\bf S}$ ($M\times M$ matrices) and 
${\bf f}$ ($M$-dimensional vector) are defined as follows:
\begin{eqnarray}
{\bf H}_{pq} &=& \langle\Phi({\bf Z}_p)|H|\Phi({\bf Z}_q)\rangle \;, \label{hpq} \\
{\bf S}_{pq} &=& \langle\Phi({\bf Z}_p)|\Phi({\bf Z}_q)\rangle \;, \label{npq} \\
{\bf f}_p &=& f({\bf Z}_p) \;. \label{fp}
\end{eqnarray}
Using this form of discretization, the optimal form of wave function (we assume the ground-state wave function in this paper) is given by the expansion
\begin{equation}
|\Psi_{GCM}\rangle   \simeq \sum_{p=1}^M {\bf f}_p e^{\Gamma({\bf Z}_p)}|\Phi\rangle
\label{gcmdiscr}
\end{equation}
which is reminiscent of recently discussed non-orthogonal variational approaches discussed in the context of quantum computing.\cite{huggins2020non} If $M$ is equal for a given basis set to a dimension of the full configuration interaction (FCI) problem (for a given spin and spatial symmetry) and $e^{\Gamma({\bf Z}_p)} |\Phi\rangle$ are linearly independent, then the expansion (\ref{gcmdiscr}) with optimized $f_i$ coefficients describes the exact electronic wave function.


\section{Multi-product extension of the Generator Coordinate formalism}

A possible extension of the GCM expansion given by Eq.(\ref{gcmexp})
can be provided by the expansion involving multiple products 
(the product GCM formalism, abbreviated as PGCM$^{(k)}$)
of $k$ exponential operator (which is inspired by a recent progress in the development of dynamical GCM methods \cite{goeke1978consistent,reinhard1978reality,reinhard1978concept,goeke1980generator,hizawa2021generator}). 
For example, one can introduce the following expansion
\begin{widetext}
\begin{equation}
|\Psi_{\rm PGCM}^{(k)}\rangle = \int d{\bf Z}(1) \ldots d{\bf Z}(k) 
 |\Phi^{(k)}({\bf Z}(1),\ldots,{\bf Z}(k))\rangle
 f({\bf Z}(1),\ldots,{\bf Z}(k))
\label{pgcmexp}
\end{equation}
\end{widetext}
where 
\begin{equation}
|\Phi^{(k)}({\bf Z}(1),\ldots,{\bf Z}(k))\rangle = e^{\Gamma({\bf Z}(k))}\ldots 
e^{\Gamma({\bf Z}(1))} |\Phi\rangle \;.
\label{pgcm}
\end{equation}
In the above representation, $\Gamma({\bf Z}(i))$ can belong to various Lie algebras. In the following we will focus on the case where all $\Gamma({\bf Z}(i))$'s
($i=1,\ldots,k)$ belong to the same $U(N)$ Lie algebra; that is,
\begin{equation}
    \Gamma({\bf Z}(i))=\sum_{pq} z_{pq}(i) E^p_q \;,\; z_{qp}(i)^{\star}=-z_{pq}(i)
\label{nun}
\end{equation}
In fact, the PGCM formula may be viewed as a special case of the GCM where 
\begin{equation}
 {\bf q} \rightarrow {\bf Z}(1)\times \ldots \times {\bf Z}(k) \;.
 \label{nun1}
\end{equation}

We will demonstrate that the $U(N)$ case of PGCM lends itself for an efficient way of representing higher-rank excitations in quantum computing. In particular, this goal can be achieved by using a simple algorithm for discretization of 
${\bf Z}(i)$ domains in the GCM method. In particular, we can show that PGCM$^{(k)}$ can be used to approximate $2k$-tuple excitations in the configuration-interaction-type expansion. 

Let us focus on the specific case when $k=2$ (i.e., the PGCM$^{(2)}$formalism). In this case we will represent the $|\Psi^{(2)}_{\rm PGCM}\rangle$ wave function, given by the formulas: 
\begin{widetext}
\begin{equation}
|\Psi_{\rm PGFM}^{(2)}\rangle = \int d{\bf Z}(1)  d{\bf Z}(2) 
 |\Phi^{(2)}({\bf Z}(1),{\bf Z}(2))\rangle
f({\bf Z}(1),{\bf Z}(2))
\label{pgcmexp2}
\end{equation}
\end{widetext}
and 
\begin{equation}
|\Phi^{(2)}({\bf Z}(1),{\bf Z}(2))\rangle = e^{\Gamma({\bf Z}(2))}
e^{\Gamma({\bf Z}(1))} |\Phi\rangle \;. \label{pgcm2}
\end{equation}
The GCM algorithm can be adapted easily for the product representation of the trial wave functions. Now the discretization of the problem, in analogy to Eq.(\ref{eigendisc}), involves a $Q$ set defined as $Q=\lbrace {\bf Z}_I \rbrace_I$, where ${\bf Z}_I = {\bf Z}(1)_p \times {\bf  Z}(2)_q$, where we use composite index $I=(p,q)$.
Controllable sampling algorithms of the sub-spaces of the Hilbert space corresponding to higher-rank excitations with the PGCM formalism based on the $U(N)$ algebras require careful selection of the sampling points, which will be discussed in the following section in the context of PGCM$^{(2)}$ method applications to the H4 system.

\section{H4 model: the choice of the GCM sampling points}\label{sec:h4model}

\begin{figure*}
    \centering
    \includegraphics[width=0.4\linewidth]{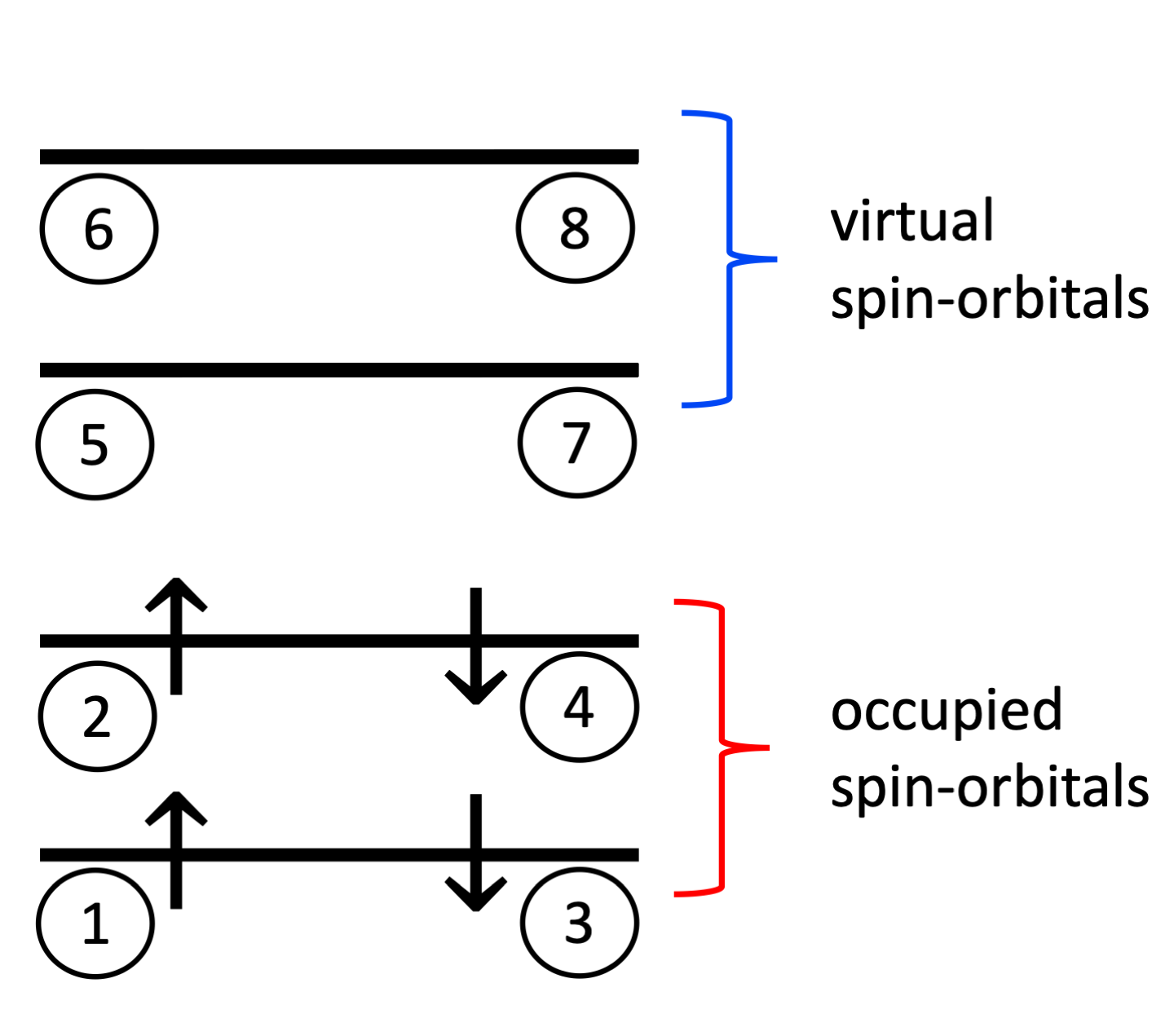}
    \caption{Schematic representation of the orbital energies and the enumeration scheme for the corresponding spin-orbitals for the H4 model in STO-3G basis set.\cite{hehre1969self} Spin-orbitals 1,2 and 3,4 correspond to occupied $\alpha$ and $\beta$ spin-orbitals, respectively, while spin-orbitals 5,6 and 7,8 correspond to virtual $\alpha$ and $\beta$ spin-orbitals.}
    \label{fig1}
\end{figure*}
We use the H4 model of Ref.\onlinecite{jankowski1980applicability} as a benchmark system. The system's geometry can be defined by a single parameter $\alpha$. For $\alpha=0.5$, the H4 model corresponds to a linear chain of hydrogen atoms with distances between adjacent hydrogen atoms equal to 2.0 a.u.; for $\alpha=0.005$, the system is almost in a square configuration. The main difference between $\alpha=0.5$ and $\alpha=0.005$ H4 models is in the structure of the corresponding ground-state wave function. While for $\alpha=0.5$, the ground-state wave function is dominated by the restricted Hartree-Fock (RHF) determinant, for $\alpha=0.005$, the ground state is quasi-degenerate. The contribution of the RHF Slater determinant is almost the same as the doubly excited configuration where two electrons from the highest occupied orbital are promoted to the lowest lying virtual orbital. The spin-orbital numbering scheme is shown in Fig.\ref{fig1}.
In the PGCM$^{(2)}$ formalism for H4, we will adopt the following sampling scheme, where the choice of the cluster operators is consistent with analyzing HF equation stability conditions based on the Thouless theorem \cite{thouless1960stability,vcivzek1967stability,seeger1977self,fukutome1981unrestricted}:
\begin{widetext}
\begin{eqnarray}
 |R_0\rangle &=& |\Phi\rangle \;, \label{x1} \\
 |R_1^{(\pm)}(t_1)\rangle &=& e^{\pm t_1 R_1} |\Phi\rangle 
 =e^{\pm t_1 \lbrace (a_5^{\dagger} a_2 + a_7^{\dagger} a_4) - 
 (a_2^{\dagger} a_5 + a_4^{\dagger} a_7) \rbrace} |\Phi\rangle
 \;, \label{x2} \\
 |R_2^{(\pm)}(t_2)\rangle &=& e^{\pm t_2 R_2} |\Phi\rangle 
 =e^{\pm t_2 \lbrace (a_6^{\dagger} a_1 + a_8^{\dagger} a_3) - 
 (a_1^{\dagger} a_6 + a_3^{\dagger} a_8) \rbrace} |\Phi\rangle
 \;, \label{x3} \\
 |R_3^{(\pm)}(t_3)\rangle &=& e^{\pm t_3 R_3} |\Phi\rangle 
 =e^{\pm t_3 \lbrace (a_6^{\dagger} a_2 + a_8^{\dagger} a_4) - 
 (a_2^{\dagger} a_6 + a_4^{\dagger} a_8) \rbrace} |\Phi\rangle
 \;, \label{x4} \\
 |R_4^{(\pm)}(t_4)\rangle &=& e^{\pm t_4 R_4} |\Phi\rangle 
 =e^{\pm t_4 \lbrace (a_5^{\dagger} a_1 + a_7^{\dagger} a_3) - 
 (a_1^{\dagger} a_5 + a_3^{\dagger} a_7) \rbrace} |\Phi\rangle
 \;, \label{x5} \\
|R_5 (t_5) \rangle &=& e^{t_5 R_3} e^{t_5 R_4} |\Phi\rangle  \;, \label{x6} \\
|R_6 (t_6) \rangle &=& e^{t_6 R_4} e^{t_6 R_3} |\Phi\rangle\;, \label{x7} \\
|R_2^{(\pm)} R_1^{(\pm)}(t_7)\rangle &=& 
e^{\pm t_7R_2} e^{\pm t_7 R_1} |\Phi\rangle
=e^{\pm t_7 \lbrace (a_6^{\dagger} a_1 + a_8^{\dagger} a_3) - 
 (a_1^{\dagger} a_6 + a_3^{\dagger} a_8) \rbrace} 
 e^{\pm t_7 \lbrace (a_5^{\dagger} a_2 + a_7^{\dagger} a_4) - 
 (a_2^{\dagger} a_5 + a_4^{\dagger} a_7) \rbrace} |\Phi\rangle \;.
 \label{x8}
\end{eqnarray}
\end{widetext}
These sampling vectors can be naturally tied to the general form of the $|\Phi^{(2)}({\bf Z}(1),{\bf Z}(2))\rangle$ basis given by Eq.(\ref{pgcm2}). For example, $|R_0\rangle$ corresponds to ${\bf Z}(1)={\bf Z}(2)=0$. For $|R_1^{(+)}(t_1)\rangle$, ${\bf Z}(2)=0$, and
$z(1)_{52}=z(1)_{74}=-z(1)_{25}=-z(1)_{47}=t_1$ while remaining matrix elements are equal to zero, etc.
Another advantage of using this form of sampling vectors is that their combinations provide a rudimentary (yet not exact) mechanism for eliminating symmetry impurities when $R_i$ operators break the symmetry of the reference state $|\Phi\rangle$.
For example, if in general $R_1^{(\pm)}(t_1)$ the operator is expressed in terms of excitations that produce a triplet state when acting on the reference function, then this singly-excited impurity (or instability) is eliminated by taking a combination of $|R_1^{(+)}\rangle + |R_1^{(-)}\rangle$ states; that is,
\begin{eqnarray}
 &&|R_1^{(+)}(t_1)\rangle + |R_1^{(-)}(t_1)\rangle =
 (e^{t_1 R_1}+e^{-t_1 R_1})|\Phi\rangle  \nonumber \\
 &&= (2+t_1^2 R_1^2 + \frac{2}{4!} t_1^4 R_1^4 + \ldots)|\Phi\rangle \;,
 \label{x9}
\end{eqnarray}
where linear triplet ``impurities'' are eliminated. 
These combinations also can allow us to selectively approach doubly excited configurations, or using combinations of $|R_2^{(\pm)} R_1^{(\pm)}(t_7)\rangle$, quadruply excited ones using manifold of single excitations and very simple quantum circuits to represent $e^{\pm tR_i}$ operators. 
This analysis can be extended to higher-order PGCM$^{(k)}$ formulations to 
include higher-rank excitations. It should be stressed that $t_i$ parameters in Eqs.(\ref{x1})-(\ref{x8}) can be chosen at random. 
The effects of random $t_i$ parameters on ground-state FCI energies are illustrated in Appendix~\ref{apdx:impl}.

\section{General outline of quantum algorithm and post-processing on conventional computers}

After selecting GCM sampling points, the remaining work is involves computing matrices ${\bf H}$ and ${\bf S}$ through Eq.(\ref{hpq}) and (\ref{npq}) using quantum computers. 
To accommodate gate-based quantum computers, the expectations are computed in the form of 
\begin{align}
    {\bf H}_{pq} &= \sum_j h_j \langle \Phi | \left(e^{-\Gamma ({\bf Z}_p)} P_j e^{\Gamma ({\bf Z}_q)} \right) |\Phi \rangle           \label{eq:realh} \\
    \text{and } {\bf S}_{pq} &= \langle \Phi | \left(e^{-\Gamma ({\bf Z}_p)} e^{\Gamma ({\bf Z}_q)} \right) |\Phi \rangle,   \label{eq:reals}
\end{align}
where Hamiltonian matrix $H$ is transformed to the linear combination of Pauli strings $H = \sum_j h_jP_j$ under Jordan-Wigner (JW) transformation and $|\Phi \rangle$ is the HF state. 
The expectations can be easily evaluated using algorithms like the Hadamard test on fault-tolerant quantum computers. A pure quantum algorithm proposed for GCM and its complexity analysis is given in Appendix \ref{LinSysQAlg}, where we propose to exploit a block encoding for the operation $\mathbf{S}^{-1}$ and use the phase estimation to compute the eigenvalues for the non-orthogonal eigenproblem (\ref{eigendisc}). Note that the pure quantum algorithm gives a favorable scaling but relies on the use of the controlled unitary circuits that makes the quantum simulations on NISQ devices challenging.

\begin{figure}[H]
    \centering
    \includegraphics[width=0.99\linewidth]{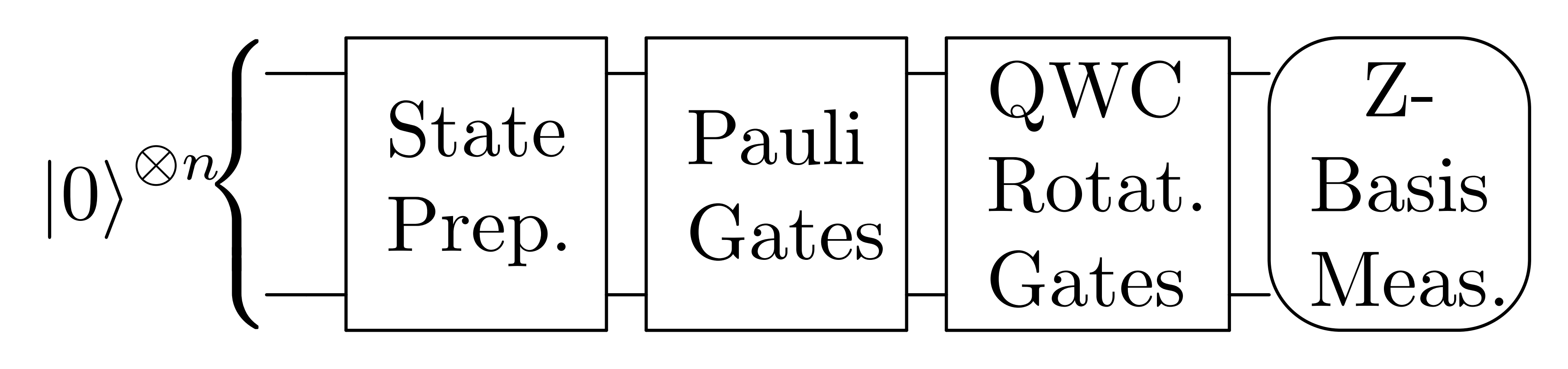}
    \caption{Shallow circuit for Alg.\ref{alg:hgcm}. State-preparation gates initialize the ground state to the HF state. Pauli gates are a single layer of 1-qubit Pauli gates. QWC rotation gates consist of 1-qubit rotation gates that allow us to measure each wire on Pauli-$X, Y,$ or $Z$ bases with the final Pauli-$Z$-basis measurement. So all gates in the circuit are 1-qubit gates and the total number of 1-qubit gates is $O(n)$.}
    \label{fig:hgcm_circ}
\end{figure}

For near-term devices, we focus on hybrid quantum-classical approach. there is a trade-off between the depth of the quantum circuit and classical computation time. 
Note that, following the JW transformation, each of the operators $R_1$ to $R_4$ in Eq.(\ref{x2}) to (\ref{x5}) can be transformed to four commuting Pauli strings.
That is to say, the matrix exponential $e^{\pm t_i R_i}$ for $i \in \{1,2,3,4\}$ are exactly the linear combinations of Pauli strings. 
So, for both H4 models, the calculations of expectation in Eq.(\ref{eq:realh}) and (\ref{eq:reals}) are essentially equivalent to $\langle \Phi |  P|\Phi \rangle $ for some Pauli string $P$, while qubit-wise commuting (QWC) terms are grouped to perform a simultaneous measurement. 
The only implemented form of the circuit is illustrated in Fig.\ref{fig:hgcm_circ} and the whole process is summarized in Alg.\ref{alg:hgcm}.
The entire simulated quantum computation is achieved using Qiskit.\cite{Qiskit} We put a brief discussion about trotterization under Qiskit in Appendix~\ref{apdx:pauli-trotter}, along with other implementation details related to duplicated Pauli strings and effects of random parameters in Appendix~\ref{apdx:impl}.

\begin{algorithm}[H]
  {\scriptsize
  \caption{Quantum GCM (QuGCM) for near-term devices  \label{alg:hgcm}}
   \begin{algorithmic}[1]
   \Require Hamiltonian matrix $H= \sum_j h_j P_j$, HF state $|\Phi\rangle$, and a set $\{\Gamma ({\bf Z}_i)\}_{i=1}^{M}$ where the index $i$ could be a composite up to $k$ terms as in Eq.(\ref{pgcm2})
   \State Transform all $\{\Gamma ({\bf Z}_i)\}_{i=1}^{M}$  using JW transformation
   \State Generate unitaries $\{V_i\}_{i = 1}^M$ for  $V_i := e^{\Gamma ({\bf Z}_i)}$ with Eq.(\ref{expz}) and Eq.(\ref{nun})
   \State Trotterize each element in $\{V_i\}_{i = 1}^M$ to a linear combination of Pauli strings \label{line:trotter}
   \For{each $V_q$ in $\{V_i\}_{i = 1}^M$} \label{line:for1}
       \State Compute $  \sum_j h_j P_j V_q$ classically \label{line:sandwichH1}
       \For{each $V_p$ in $\{V_i\}_{i = 1}^M$} \label{line:for2}
            \State Compute $  \sum_j h_j V_p^{\dagger} (P_j V_q)$ classically \label{line:sandwichH2}
            \State Compute $V_p^{\dagger} V_q$ classically \label{line:sandwichS}
            \State Evaluate ${\bf H}_{pq} := \sum_j h_j \langle \Phi |V_p^{\dagger} P_j V_q| \Phi \rangle$ in a quantum device \label{line:computeH}
            \State Evaluate ${\bf S}_{pq} := \langle \Phi |V_p^{\dagger} V_q| \Phi \rangle$ in a quantum device \label{line:computeS}
       \EndFor
   \EndFor
   \State Solve the general eigenvalue problem ${\bf H} {\bf f} =E {\bf S} {\bf f}$ classically
   \State \Return interested eigenvalues and eigenvectors
   \end{algorithmic}
   }
\end{algorithm}

\section{Complexity Analysis and effects of finite samplings}



Recall that $M$ is the number of selected ${\bf Z}$-points, which is equivalent to the number of sampling vectors. 
Let $n$ be the number of spin orbitals.
We know the Hamiltonian matrix, $H$, consist of at most $O(n^4)$ Pauli strings, so operations for multiplying matrices and solving the generalized eigenvalue problem in Alg.\ref{alg:hgcm} dominate the classical part of the operation complexity.
If an exponent of a cluster operator has up to $c$ number of Pauli strings after trotterization and the multiplication between two length-$n$ Pauli strings takes at most $O(n)$ operations, then the matrix multiplication in line \ref{line:computeH} and \ref{line:computeS} of Alg.\ref{alg:hgcm} takes $O(n \cdot c^2 \cdot n^4)$ and $O(n \cdot c^2)$ operations, respectively, because all matrices are decomposed into the linear combinations of Pauli terms.
As we have $M^2$ iterations in Alg.\ref{alg:hgcm} and the final generalized eigenvalue problem of $M \times M$ matrices takes $O(M^3)$ operations, Hybrid PGCM$^{(2)}$ has the overall classical part of the worst-case scaling of $O(c^2 n^5  M^2 + M^3)$ operations.

The specific value of $c$ highly depends on the how users want to approximate the molecular models. 
It is affected by the number of creation and annihilation operators pairs in each single-excitation cluster operators, the value of $k$, and number of trotterization steps if it is necessary. 
As we discussed earlier, we can properly choose two pairs of creation and annihilation operators in each of single-excitation cluster operators without the requirements of trotterization.
The results in Section \ref{sec:results} are promising enough when only single and double excitations are considered, i.e., when $k=2$.

Regarding the quantum part of the operation complexity, it includes the number of circuits and the number of measurements in every circuit. Naively, line \ref{line:computeH} and line \ref{line:computeS} in Alg.\ref{alg:hgcm} indicate there are $O(n \cdot c^2 \cdot n^4)$ number of circuits and every circuit contains only $O(n)$ number of 1-qubit gates. 
Many existing methods can be applied to reduce the number of terms associated with $V_p^{\dagger} (H V_q)$ and $V_p^{\dagger} V_q$. 
For example, to reduce the order $O(n \cdot c^2 \cdot n^4)$ to $O(n \cdot c^2 \cdot n^{2 \sim 3})$ for the number of circuits in every iteration, the linear combination of unitaries technique,\cite{Childs2012} amplitude amplification approach,\cite{brassard2002quantum} Hamiltonian simulation,\cite{Berry2015Simulating} qubitization,\cite{low2017optimal} or the direct block-encoding\cite{gilyen2019quantum} methods can be typically employed at the cost of introducing deeper circuits and implementing controlled unitary operations.
Nevertheless, these approaches come with a probability of failure and require advanced circuit and error mitigation that might go beyond the capability of the current NISQ devices.
Toward a more feasible NISQ approach, in light of the unitary partitioning scheme proposed by Izmaylov et al.\cite{Izmaylov2019}, Peng and Kowalski recently proposed a more efficient unitary partitioning approach guided by the single-reference trial state used in the simulation.\cite{peng2022mapping}
In particular, through numerical tests over a wide range of molecules in different bases, they found that the non-unitary wave operators when acting on single-reference trial wave function (such as $e^{\Gamma(\mathbf{Z}_i)}|\Phi\rangle$ and $He^{\Gamma(\mathbf{Z}_i)}|\Phi\rangle$ in the present discussion) can be represented by a much more compact unitary basis, thus providing a more efficient route for performing the general non-unitary quantum simulations.

It is worth mentioning that the above discussion is focused on the number of terms/operations that can be efficiently reduced through groupings that feature the commutativity or anti-commutativity of Pauli strings, while the total number of measurements required from the number of groups also critically depends on the covariance between the contributing terms, $\text{cov}\big(P_i,P_j\big)$, and the desired precision $\epsilon$. Given that we can always write a matrix operator as a linear combination of Pauli strings, the total number of measurements for evaluating the expectation value of an operator with respect to the trial wave function can then be expressed as \cite{gonthier2020identifying,Rubin2018Hybrid,PhysRevA.92.042303}
\begin{align}
  & \text{\# of Measurements} = \notag \\
   &~~~~~~~~~~~~~~~~~~~~\Bigg( \displaystyle\frac{\sum_G\sqrt{\sum_{i,j,\in G}h_i h_j \text{cov}\big(P_i,P_j\big)}}{\epsilon}\Bigg)^2
   \label{eq:Measurement}
\end{align}
with $G$ indexing the groups. Therefore, it is likely that the number of groups decreases at the cost of introducing larger covariances that could essentially increase the total number of measurements required to achieve a desired precision. 
The variance of an individual Pauli string $P_i$ can be bounded by
\begin{eqnarray}
    \text{var}(P_i) = \braket{P_i^2} - \braket{P_i}^2 = 1 - \braket{P_i}^2 \leq 1,
\end{eqnarray}
which provides bounds to the covariance of any contributing terms
\begin{eqnarray}
    \left| \text{cov}(P_i, P_j) \right| \leq \text{var}(P_i)\text{var}(P_j) \leq 1.
\end{eqnarray}
Thus, according to Eq.(\ref{eq:Measurement}), the bounds for the standard deviations of the matrix entry ${\bf H}_{pq}$ and ${\bf S}_{pq}$ under a finite number of measurements are
\begin{eqnarray}
    \epsilon_{{\bf H}_{pq}} &\leq \frac{\sum_{G_{{\bf H}_{pq}}}\sqrt{\sum_{i,j,\in G_{{\bf H}_{pq}}} |h_i h_j|}}{\sqrt{\text{\# of Measurements}}}  \label{eq:max_shot_var_h} \\
    \epsilon_{{\bf S}_{pq}} &\leq \frac{\sum_{G_{{\bf S}_{pq}}}\sqrt{\sum_{i,j,\in G_{{\bf S}_{pq}}} |h_i h_j|}}{\sqrt{\text{\# of Measurements}}} \label{eq:max_shot_var_s}
\end{eqnarray}

\begin{figure}[h]
    \centering
    \includegraphics[width=0.99\linewidth]{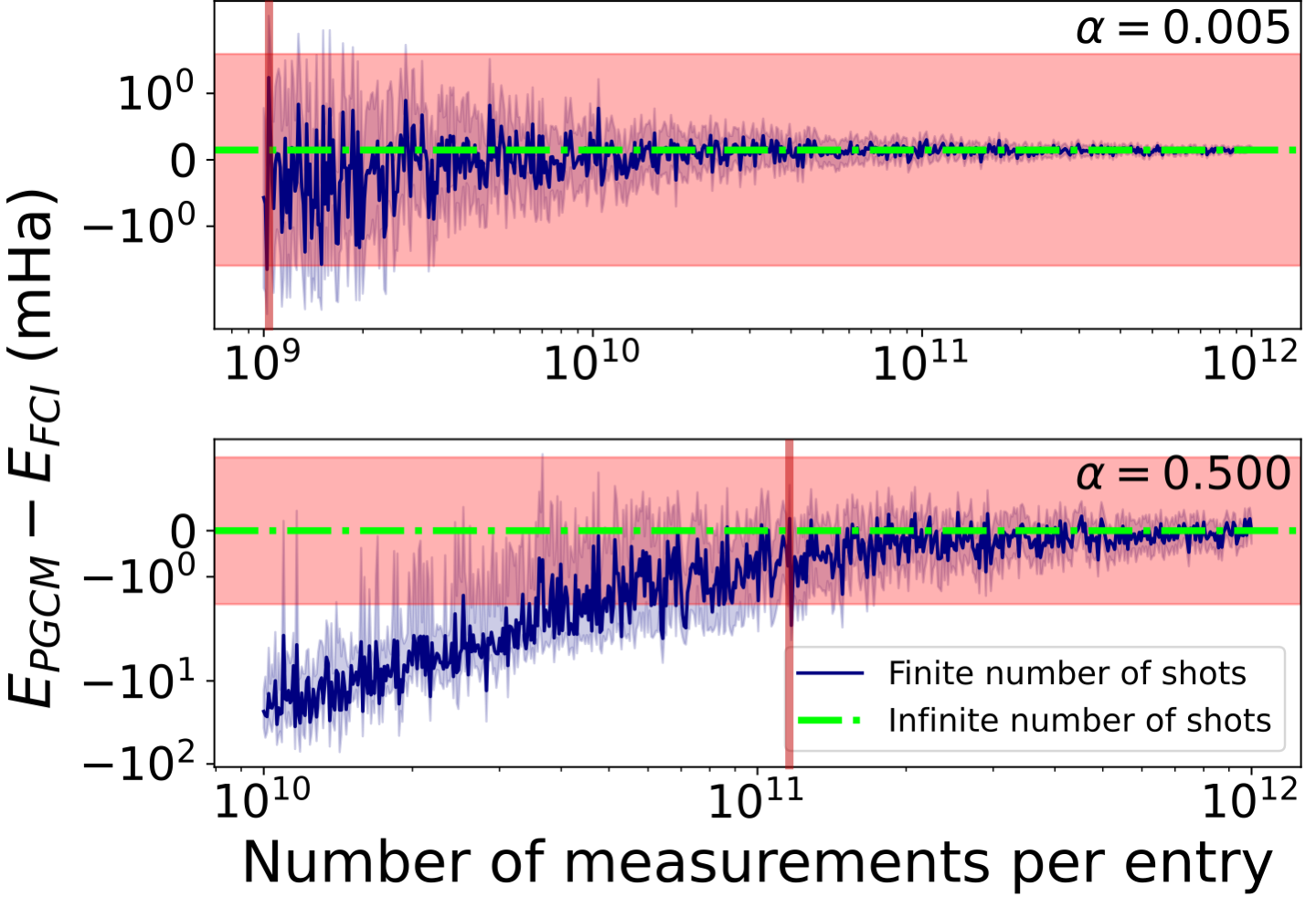}
    \caption{Differences between the ground-state energy estimations from PGCM$^{(2)}$ and FCI formalism in milli-Hartrees for $\alpha = 0.005$ (up) and $\alpha=0.500$ (down) H4 models, respectively. The variance of each random sampling is set to the maximum according to Eq.(\ref{eq:max_shot_var_h}) and Eq.(\ref{eq:max_shot_var_s}). The red-shaded region is the range of chemical accuracy $\pm 1.5936 \text{mHa}$. The blue-shaded region is the $95\%$ confidence interval estimated from 100 simulations with the red bars marking the rough number of measurement per entry at which the energy difference would be within the chemical accuracy.}
    \label{fig:per}
\end{figure}

Let $\tilde{{\bf H}}_{pq}$ and $\tilde{{\bf S}}_{pq}$ be the entries computed from finite number of measurements. We can empirically estimate the effects of the uncertainty by considering $\tilde{{\bf H}}_{pq}$ and $\tilde{{\bf S}}_{pq}$ as normal random variables
\begin{eqnarray}
    \tilde{{\bf H}}_{pq} &\sim N \left({\bf H}_{pq},\epsilon^2_{{\bf H}_{pq}} \right) \text{ and }
    \tilde{{\bf S}}_{pq} &\sim N \left({\bf S}_{pq},\epsilon^2_{{\bf S}_{pq}} \right).
\end{eqnarray}
To illustrate the maximum possible fluctuations brought by finite-sampling errors, we set each $\epsilon_{{\bf H}_{pq}}$ and $\epsilon_{{\bf S}_{pq}}$ to their maximum as in Eq.(\ref{eq:max_shot_var_h}) and Eq.(\ref{eq:max_shot_var_s}), and summarize the results in Fig.\ref{fig:per}. It is clear that to generally reach chemical accuracy, the $\alpha = 0.005$ case requires about two orders of magnitude more measurements than the $\alpha = 0.500$ case using Alg.\ref{alg:hgcm}.

We also notice that there are also many other advanced measurement schemes proposed recently. One example is to simultaneously obtain expectation values of multiple observations by randomly measuring and projecting the quantum state into classical shadows \cite{huang2020predicting,huang2021efficient,aaronson2018shadow,struchalin2021experimental,chen2021robust,zhao2021fermionic,acharya2021informationally,hadfield2021adaptive,hillmich2021decision,zhang2021experimental}. In principle, the algorithm enables measurements of $m$ low-weight observations using only $O(\log_2m)$ samples. The practical performance of the algorithm for model and molecular Hamiltonians on NISQ devices, in terms of accuracy and efficiency, is still under intense study.

\section{Results \label{sec:results}}

The GCM results for the ground-state energies and excitation energies corresponding to low-lying states of the symmetry of the reference function (singlet $A_1$ states) are collated in Tables \ref{tab:ground} and \ref{tab:esinglets}, respectively. To evaluate the accuracy of ground-state simulations, we compared GCM results with the results obtained with the RHF, multi-configurational self-consistent field (MCSCF) formalism for active space defined by four electrons and three active orbitals (MCSCF(4e,3o)), configuration interaction method with singles and doubles (CISD), CC method with singles and doubles (CCSD), VQE formalism, and FCI formalism. We used the equation-of-motion CC approach (EOMCCSD) and FCI formalism for excited states. Note that any quantum computation in VQE and GCM assume infinite number of measurements.

Inspection of the results in Table \ref{tab:ground} indicates that for both geometries, the GCM results are in very good agreement with the FCI result despite the simplicity of the expansions given by Eqs.(\ref{x1})-(\ref{x8}). Interestingly, the GCM energies are significantly better in both cases than the VQE ones. The excellent performance of the GCM formalism is well illustrated by the strongly correlated $\alpha=0.005$ case, where CISD and CCSD methods struggle to capture needed correlation effects. For the weakly correlated variant of H4 ($\alpha=0.500$), the GCM formalism reproduces nearly FCI-level accuracy with the energy error of 0.022 milli-Hartree.

In Table \ref{tab:esinglets} we juxtaposed the excitation energies ($\omega_1$, $\omega_2$, and $\omega_3$) obtained with the GCM approaches for three lowest-lying $^1A_1$ symmetry states, of H4 model for $\alpha=0.005$ and $\alpha=0.500$. For the $\omega_1$ excitation energies, the GCM approach provides consistently better estimates of their exact (FCI) values than the ubiquitous EOMCCSD approach. While for $\omega_2$ GCM prediction is better than the EOMCCSD one only for the $\alpha=0.005$, for non-degenerate case ($\alpha=0.500$), the GCM prediction is by 0.6 eV off the FCI value. For the $\omega_3$ excitation energies, the GCM is capable of providing estimates within 0.160 eV ($\alpha=0.005$) and 0.33 eV ($\alpha=0.500$) of error. Again, given the simplicity of the GCM formulations, one should view the GCM estimates of the excitation energies as quite satisfactory. 
\renewcommand{\tabcolsep}{0.15cm}
\begin{table*}[!ht]
    \centering
    \begin{tabular}{c c c c c c c }
\hline\hline \\[-0.2cm]
        FCI & RHF  & MCSCF(4e,3o) & CISD  & CCSD & VQE & GCM  \\[0.1cm]
        \hline \\[-0.2cm]
        \multicolumn{7}{c}{H4 $\alpha=0.005$} \\[0.1cm]
        \hline  \\[-0.2cm]
        -1.942993 & 151.407 & 82.645   & 5.507 & 3.331 & 0.905 &  0.147 \\[0.1cm]
        \hline \\[-0.2cm]
         \multicolumn{7}{c}{H4 $\alpha=0.500$} \\[0.1cm]
         \hline \\[-0.2cm]
        -2.151007 &  75.764 &  43.276  & 1.866 &  0.003 & 0.058 & 0.022 \\[0.1cm]
         \hline
    \end{tabular}
    \caption{Differences (in milliHartrees) with total ground-state  FCI energies.}
    \label{tab:ground}
\end{table*}
%
%
%
\renewcommand{\tabcolsep}{0.45cm}
\begin{table*}[!ht]
    \centering
    \begin{tabular}{l c c c}
\hline\hline \\[-0.2cm]
        Method & $\omega_1$ & $\omega_2$ & $\omega_3$ \\[0.1cm]
        \hline \\[-0.2cm]
        \multicolumn{4}{c}{H4 $\alpha=0.005$} \\[0.1cm]
        \hline  \\[-0.2cm]
        FCI          & 4.183 & 6.040 & 18.484  \\[0.1cm]
        EOMCCSD      & 4.275 & 6.079 & 18.350  \\[0.1cm]
        GCM         & 4.179 &  6.038  & 18.635 \\[0.1cm] 
        \hline \\[-0.2cm]
         \multicolumn{4}{c}{H4 $\alpha=0.500$} \\[0.1cm]
         \hline \\[-0.2cm]
        FCI          & 12.565 & 14.214 & 21.293 \\[0.1cm]
        EOMCCSD    & 12.738 & 14.179 & 21.255 \\[0.1cm]
        GCM     & 12.589  & 14.840 & 21.622 \\[0.1cm]
         \hline
    \end{tabular}
    \caption{Singlet excitation energies in eV}
    \label{tab:esinglets}
\end{table*}
%

\section{Conclusion}
In this work, we explored use of the GCM in the context of quantum computing. For this purpose, we introduced the multi-product extension of the GCM formalism that enables one to construct state vectors in Hilbert space using various types of the Fermion Lie algebra and general quantum algorithms that allow one to perform GCM calculations in a way that can be viewed as a specific case of the quantum algorithms for configuration interaction formalisms involving non-orthogonal basis.
In the present studies, we focused entirely on the $U(N)$ algebra, where resulting state vectors can be interpreted in terms of the Thouless theorem. This analogy is essential in the sampling process of the parametrized unitary canonical transformations. It enables one to construct
corresponding state vectors and corresponding linear space where higher-order excitations (e.g., double, triple, quadruple, etc.) excitations
can be selectively approached using the language of single excitations to define generators ($\Gamma({\bf Z})$) of the canonical transformations. The discussed procedures can be easily related to the searches of various type instabilities in independent particle formulations with the HF method as a specific example.

Using the H4 system as a benchmark, we showed that the quantum GCM algorithm could provide ground-state energies competitive to the VQE simulations involving the explicit form of the double excitations. In contrast to standard VQE algorithms, the GCM formalism also yields the values of excited-state energies. We showed that GCM excitation energies corresponding to the low-lying excited states could be competitive with the excitation energies obtained with the popular EOMCCSD approach.

Although in the present form, the GCM formalism falls into the category of hybrid formulations, unlike the VQE method, it avoids multiple quantum-classical machine data transfer and unstable and tedious iterative processes. Also, our near-term implementation of the quantum GCM algorithm can run in parallel easily using multiple quantum and classical nodes.

In future studies, we plan to use the full potential of the quantum algorithms based on the quantum CI formulations and explore the possibility of using other types of Fermionic algebras, i.e., $SO(2N)$ and $SO(2N+1)$ Lie algebras.

\section{ACKNOWLEDGMENTS}
Muqing Zheng, Bo Peng, Nathan Wiebe, Ang Li, and Karol Kowalski were supported by  the ``Embedding QC into Many-body Frameworks for Strongly Correlated Molecular and Materials Systems'' project, which is funded by the U.S. Department of Energy, Office of Science, Office of Basic Energy Sciences, the Division of Chemical Sciences, Geosciences, and Biosciences. The Pacific Northwest National Laboratory is operated by Battelle for the U.S. Department of Energy under Contract DE-AC05-76RL01830. Xiu Yang was supported by National Science Foundation CAREER DMS-2143915. Muqing Zheng and Xiu Yang both also were supported by Defense Advanced Research Projects Agency as part of the project W911NF2010022: {\em The Quantum
Computing Revolution and Optimization: Challenges and Opportunities}.

\section{Data availability}
The code and data in the paper are openly available in the GitHub repository.\cite{data_repo}








\appendix

\section[\appendixname~\thesection]{Duplicated Pauli strings and selection of parameters \label{apdx:impl}}

For each of H4 models, while $M = 15$, the symmetries of ${\bf H}$ and ${\bf S}$ allows us to only compute 120 iterations instead of 225 iterations. Among those 120 iterations, as shown in Table \ref{tab:duppauli}, we only need to measure 4,000 Pauli strings because more than $97\%$ of them are duplicated. This brings the two orders of magnitude reduction on the total number of measurements in practice. 

\renewcommand{\tabcolsep}{0.1cm}
\begin{table}
    \centering
    \begin{tabular}{c c c c}
\hline\hline \\[-0.2cm]
          & Num. of unique terms & Total num. of terms  & Ratio ($\%$)  \\[0.1cm]
        \hline \\[-0.2cm]
        \multicolumn{4}{c}{H4 $\alpha=0.005$} \\[0.1cm]
        \hline  \\[-0.2cm]
        ${\bf S}$  &$1444$  &$38904$   &$3.71\%$\\[0.1cm]
        ${\bf H}$  &$3423$  &$118426$   &$2.89\%$\\[0.1cm]
        All  &$4180$  &$157330$   &$2.66\%$\\[0.1cm]
        \hline \\[-0.2cm]
         \multicolumn{4}{c}{H4 $\alpha=0.500$} \\[0.1cm]
         \hline \\[-0.2cm]
        ${\bf S}$  &$1766$  &$38888$   &$4.54\%$\\[0.1cm]
        ${\bf H}$  &$3454$  &$118621$   &$2.91\%$\\[0.1cm]
        All  &$4370$  &$157509$   &$2.77\%$\\[0.1cm]
         \hline
    \end{tabular}
    \caption{Number of unique Pauli terms and their ratios after grouping among all iterations.}
    \label{tab:duppauli}
\end{table}

\begin{figure}
    \centering
    \includegraphics[width=0.99\linewidth]{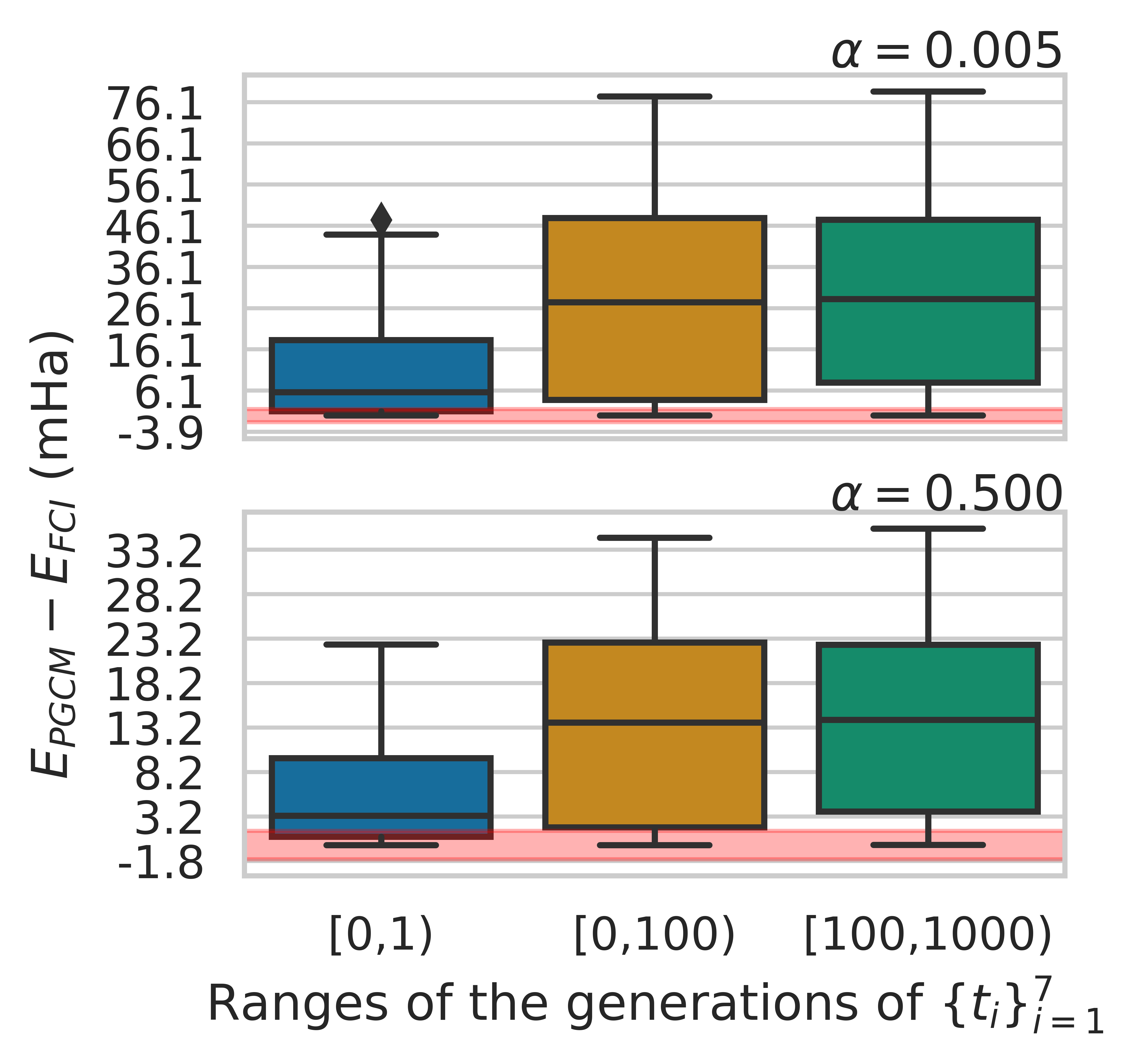}
    \caption{Differences between the ground-state energy estimations from PGCM$^{(2)}$ in various random $\{t_i\}$'s and FCI formalism in milli-Hartrees for $\alpha = 0.005$ (up) and $\alpha=0.500$ (down) H4 models, respectively. The horizontal red-shaded region shows the range of chemical accuracy $\pm 1.5936 \text{ mHa}$. The fact that minimum value and the 25$^{th}$ percentile are in the shaded region when $t_i \in [0,1)$ in both H4 models indicates that random generations of ${t_i}$ in that range is relatively likely to provide a good approximation in our algorithm.}
    \label{fig:ts}
\end{figure}

Meanwhile, we conducted the following experiments for $\alpha = 0.005$ and $\alpha=0.500$ H4 models to demonstrate the influence of the random choices of $t_i$ parameters on the estimations of the ground-state FCI energies.
For each molecular model, we generated 50 sets of $\{t_i\}_{i = 1}^7$ for $t_i$ in $[0,1)$, $[0,100)$, and $[100,100)$, respectively. Each of 300 sets of $\{t_i\}_{i = 1}^7$ produced an estimation of the ground-state FCI energy from Alg.~\ref{alg:hgcm} (assume infinite number of shots).
The distributions of differences between the estimations from PGCM$^{(2)}$ and FCI formalism under various settings are illustrated in Fig.\ref{fig:ts} using box-plots.
It worth noting that when we produce $t_i$'s from $[0,1)$, about $25\%$ random generations can provide an estimation of ground-state energy within the range of chemical accuracy.\\

\section[\appendixname~\thesection]{Linear-System Inspired Algorithms for GCM}\label{LinSysQAlg}

Perhaps the most direct way to use quantum computing to solve the electronic structure problem using Generator Coordinate Methods is by simply solving the non-orthogonal eigenvalue problem by dilating it to a square matrix in a higher dimensional space. In this case, it is most convenient to express our eigenvalue problem as
\begin{equation}
    {\bf S}^{-1}{\bf H} {\bf f}=E  {\bf f},
\end{equation}
where we have assumed here that ${\bf S}$ is an invertible matrix.
Now let us define an isometric extension of our original space. We do this to exploit a block encoding for the operation ${\bf S}^{-1}$, which allows us to express it as a unitary operation in a higher dimensional space. Specifically, let
\begin{align}
    {\bf U} &= \begin{bmatrix}{\bf S}^{-1}/\|{\bf S}^{-1}\| & {\Box} \\ {\Box} & {\Box} \end{bmatrix},
\end{align}
be a unitary matrix (i.e., ${\bf U}^\dagger ={\bf U}^{-1}$) for arbitrary matrices ${\Box}$. Also let
\begin{align}
    {\bf J} &= \begin{bmatrix}{\bf H}/\alpha & 0 \\ 0 & 0 \end{bmatrix}, ~~ {\bf g} = \begin{bmatrix} {\bf f} \\ 0 \end{bmatrix},~~\bf{Z} = \begin{bmatrix} \bf{I} & 0 \\ 0 & -\bf{I}\end{bmatrix},~~
    {\bf P} = \frac{1}{2}\left(\bf{I}+\bf{Z} \right). 
\end{align}
Inside this enlarged space, the eigenvalue equation reads for eigenvalue $E$
\begin{equation}
    {\bf P} {\bf U} {\bf J} {\bf P}  {\bf g}= \frac{E \bf{g}}{\|{\bf S}^{-1}\| \alpha}.
\end{equation}
Now let ${\bf H}/\alpha=\sum_j h_j {\bf U}^{(H)}_j$ for a set of unitary ${\bf U}_j^{(H)}$. We then have that our Hamiltonian in the enlarged space can be expressed as a similar linear combination of unitaries.
\begin{equation}
    {\bf J} =  \sum_{j} h_j \ketbra{0}{0}\otimes {\bf U}^{(H)}_j = \sum_{j} \frac{h_j}{2} \left({\bf I}\otimes {\bf U}^{(H)}_j+\mathbf{Z}\otimes {\bf U}^{(H)}_j\right),
\end{equation}
Thus the entire product can be expressed as
\begin{equation}
    {\bf P}{\bf U}{\bf J} {\bf P}= \sum_{j} \frac{h_j}{2} {\bf P}\left({\bf{U}}\left({\bf I}\otimes {\bf U}^{(H)}_j\right)+{\bf{U}}\left(\mathbf{Z}\otimes {\bf U}^{(H)}_j\right)\right){\bf P},
\end{equation}
Thus the enlarged Hamiltonian ${\bf J}$ can be expressed as a linear combination of unitary matrices that further has the exact same value of $\alpha = \sum_j |h_j|$.

For us to use phase estimation to compute these eigenvalues, we also need a further modification. Note that $e^{-i {\bf U J}}$ is not necessarily unitary because ${\bf U J}$ is not necessarily Hermitian. We address this issue by considering a further embedding:
\begin{align}
    {\bf K} &= \begin{bmatrix} 0 & {\bf PU JP} \\ {\bf PJ} {\bf U}^\dagger{\bf P} &0 \end{bmatrix} \notag \\
        &= \ketbra{0}{1} \otimes {\bf PUJP} + \ketbra{1}{0} \otimes {\bf PJ}{\bf U}^\dagger{\bf P} \nonumber\\
        &= \frac{1}{2} \left({\bf X}\otimes {\bf PUJP} +(i {\bf Y})\otimes {\bf PUJP}+{\bf X} \otimes {\bf PJ}{\bf U}^\dagger{\bf P} \right. \nonumber\\
        &~~~~~~~~~\left. +(-i {\bf Y})\otimes {\bf PJ} {\bf U}^\dagger{\bf P}\right)
\end{align}
This expression also takes the form of a linear combination of unitary matrices; however, the coefficient sum now obeys $\sum_j |h_j'| = 2\alpha$, which does not change the overall complexity. Finally note that ${\bf K}$ is anti-diagonal block matrix.
We can therefore search for an eigenvector of the form ${\bf h} = \begin{bmatrix}0 & {\bf g} \end{bmatrix}^\dagger$ as the eigenvectors of ${\bf K}$ can be taken to have support only on one of the two blocks.
The eigenvalues of $\mathbf{ K}$ corresponding the the eigenvector $\mathbf{g}$ of ${\bf PUJP}$ must be $\pm E$ from this construction. To see this note that
\begin{equation}
    {\rm Tr}(\mathbf{K}^2 \mathbf{h} \mathbf{h}^\dagger) = {\rm Tr}(\mathbf{h^\dagger K^{\mathrm 2} h}) = (\mathbf{Kh})^\dagger (\mathbf{Kh}) = E^2.
\end{equation}
Thus $\mathbf{h}$ is an eigenvector of $\mathbf{K}$ and the eigenvalues of $\mathbf{K}$ in the support of $\mathbf{h}$ must be $\pm E$. Thus, we can estimate the absolute value of $E$ by using phase estimation on ${\bf K}$.

To perform this simulation using qubitization ideas, we will need to propose \textit{prepare-and-select} circuits for the coefficients. Specifically,
\begin{equation}
    U_{\rm prep}\ket{0} = \sum_{j} \sum_{\mu=0,1} \sum_{\nu=0,1} \sum_{\omega=0,1} \frac{\sqrt{h_j/4}}{\sqrt{2 \alpha}} \ket{j} \ket{\mu} \ket{\nu} \ket{\omega}
\end{equation}
and further:
\begin{align}
U_{\rm sel} \ket{j} \ket{\mu} \ket{\nu} \ket{\omega} \ket{\psi}&=    \ket{j}\ket{\mu} \ket{\nu} \ket{\omega} \nonumber\\
&~~\otimes  (-1)^{\mu\omega}{\bf X}^{\mu} (i{\bf Y})^{1-\mu} \otimes {\bf Z}^{\nu} \nonumber\\
&~~\otimes {{\bf U}^\dagger}^{\omega} {\bf U}_j^{(H)}{\bf U}^{1-\omega}\ket{\psi} 
\end{align}
This then forms a block encoding of our operator for the generalized eigenvalue problem:
\begin{equation}
(\bra{0} \otimes \mathbf{I})U_{\rm prep}^\dagger U_{\rm sel} U_{\rm prep} (\ket{0} \otimes \mathbf{I})={\bf K}/(2\alpha)    
\end{equation}
Using qubitization, we can convert this into a unitary with result~\cite{gilyen2019quantum}, we can perform phase estimation to learn $E$ with a variance of at most $\epsilon^2$ using a number of queries to $U_{\rm sel}$ and $U_{\rm prep}$ that is in $O\left(\alpha\|{\bf S}^{-1}\|/\epsilon \right)$. However, each application of $U_{\rm sel}$ requires $O(1)$ queries to ${\bf U}$, which is a block encoding of the inverse of ${\bf S}$. A deeper analysis of the cost is possible; however, to do this, we first need to have consensus on the input model used for the simulation.

The easiest way to compute a matrix element for the overlap matrix is through the use of a controlled unitary. The overlap that, the estimate takes the form:
\begin{equation}
    \bra{\Phi}V^\dagger_p V_q\ket{\Phi} = {\bf S}_{p,q}.
\end{equation}
where $V_p$ is the basis transform operation such that for the reference state $\ket{\Phi}$, $V_p\ket{\Phi} = \ket{\Phi({\bf Z}_p)}$. 
To use the Hadamard test to reconstruct this circuit we require a single application of $V^\dagger_p$ and a single application of $V_q$. The probability that the control qubit that governs this circuit is $0$ is:: 
\begin{equation}
    P(0|p,q) = \frac{1+ {\rm Re}(\langle{\Phi({\bf Z}_p)}|{\Phi({\bf Z}_q)}\rangle)}{2}.
\end{equation}
If needed, the imaginary part can be similarly found by applying an $S^\dagger$ gate to the control.
If we apply amplitude amplification to the result, then we can construct a matrix $\mathbf{AA(P)}$ with eigenvalues inside the sector:
\begin{equation}
    \lambda(\mathbf{AA(P)}) = e^{\pm i \cos^{-1}\left( \sqrt{\frac{1+ {\rm Re}(\langle{\Phi(Z_p)}|{\Phi(Z_q)}\rangle)}{2}} \right)},
\end{equation}
and is equivalent to the following matrix in the two dimensional space spanned by the initial state and the marked state:
\begin{equation}
\begin{bmatrix}
\sqrt{\frac{1+ {\rm Re}(\langle{\Phi(Z_p)}|{\Phi(Z_q)}\rangle)}{2}} & -\sqrt{\frac{1- {\rm Re}(\langle{\Phi(Z_p)}|{\Phi(Z_q)}\rangle)}{2}} \\
 \sqrt{\frac{1- {\rm Re}(\langle{\Phi(Z_p)}|{\Phi(Z_q)}\rangle)}{2}} & \sqrt{\frac{1+ {\rm Re}(\langle{\Phi(Z_p)}|{\Phi(Z_q)}\rangle)}{2}}
\end{bmatrix}
\end{equation}
Applying quantum signal processing we can apply a transformation $u\mapsto 2u^2-1$. Note that this is 1) an even degree polynomial and 2 it is between $[-1,1]$ for $u$ in a similar range. This means that quantum signal processing can be used to apply this transformation on the top block of the matrix to prepare a block encoding of the form~\cite{gilyen2019quantum}:
\begin{equation}
 U_{\bf S}=\begin{bmatrix}
   {{{\bf Re}(\langle{\Phi(Z_p)}|{\Phi(Z_q)}\rangle)}} & \Box\\
    \Box & \Box
\end{bmatrix}
\end{equation}
Where specifically we have that, the initial state used in the amplitude amplification routine block-encodes the overlap matrix $\mathbf{S}$. In all, this process costs $O({\rm polylog}(1/\epsilon))$ queries to the state preparation oracle to produce this block encoding.

Next we need to prepare a block encoding of ${\mathbf{S}}^{-1}$. Using the results of Ref. \citenum{an2022quantum} we can prepare such a block encoding using $O(\|{\bf S}\| \|{\bf S}^{-1}\|{\rm polylog}(1/\epsilon))$ queries where $\epsilon$ is our target accuracy. Thus, the overall query complexity is the number of queries made to $\mathbf{K}$ multiplied by the number of queries per $U$ or $U_j^{(H)}$. The former result then shows that the final cost of the simulation using this approach (in terms of queries to the prepare and select oracles of ${\mathbf{H}}$ and the queries to $Z_p,Z_q$ scale as:
\begin{equation}
\tilde{O}\left(\frac{\alpha\| {\bf{S}}\| \|{\bf S}^{-1}\|^2}{\epsilon} \right).
\end{equation}

This shows that the query complexity of the simulation using this approach does not necessarily scale with the dimension of the space. It does, however, depend strongly on the one-norm of the coefficients of the Hamiltonian and the norm of the inverse of the overlap matrix. Thus, the worse conditioned the matrix is, the worse we expect the performance of the algorithm to be. In contrast, learning the matrix $\mathbf{S}$ in high dimensional spaces to perform the inverse can be expensive as noted in the main body. This approach gives a theoretical alternative in such cases that has favorable scaling asymptotically at the price of the algorithm requiring a large number of qubits.


\section[\appendixname~\thesection]{Trotterization of a matrix exponential \label{apdx:pauli-trotter}}

This section aims to transform a matrix exponential into a linear combination of Pauli group matrices.
Because the standard form of a matrix exponential in Qiskit is $e^{-itA}$ instead of $e^{tA}$ as we have in the main body for some parameter $t$ and matrix $A$ and we implemented the quantum part of our algorithm in Qiskit, we keep $i = \sqrt{-1}$ in the power. 
Because we will use dot product between matrices, we do not omit the operator $\otimes$ when we do the Kronecker product between Pauli matrices for clarity (e.g., $XY$ means the dot product of $X$ and $Y$ instead of $X \otimes Y$). 
Let $n$ be the number of qubits, $I_n$ the identity matrix in space $\mathbb{C}^{2^n} \times \mathbb{C}^{2^n}$, and $P \in \{X,Y,Z,I\}^{\otimes n}$ an $n$-qubit Pauli group matrix.
Then, for any $n$-qubit Pauli group matrix, we have:
\begin{equation}
    P^2 = \left(\bigotimes_{l = 1}^{2^n} \sigma_l \right) \left(\bigotimes_{l = 1}^{2^n} \sigma_l \right) = \left(\bigotimes_{l = 1}^{2^n} \sigma_l^2 \right) = I_n \label{eq:pauli_str_iden}
\end{equation}
given $\sigma_l^2 = I$ , for all $\sigma_l \in \{X,Y,Z,I\}$ and the property of Kronecker product
\begin{equation}
    (A \otimes B)(C \otimes D) = (AC) \otimes (BD) \label{eq:kron_dot}
\end{equation}
for some matrices $A,B,C,D$ in the appropriate dimensions.

Now, by doing Taylor expansion of $e^{itP}$ at $t=0$, we obtain the following exact conversion:
\begin{eqnarray}
    e^{itP} &=& I_n + (itP) + \frac{(itP)^2}{2!} + \frac{(itP)^3}{3!} + \frac{(itP)^4}{4!} + \frac{(itP)^5}{5!}\cdots \nonumber\\
    &=& I_n + itP - I_n\frac{t^2}{2} - iP\frac{t^3}{3!} + I_n\frac{t^4}{4} + iP \frac{t^5}{5!} \cdots \qquad  \nonumber\\
    &=& I_n \left(1 - \frac{t^2}{2} + \frac{t^4}{4} + \cdots\right) + iP\left(t - \frac{t^3}{3!} + \frac{t^5}{5!} + \cdots\right) \nonumber \\
    &=& \cos(t) I_n + i \sin(t) P.  \label{eq:exp_pauli_str}
\end{eqnarray}
where $t$ is a scalar parameter.
If we deal with more than a single Pauli group matrix, then we need to use Suzuki trotterization. For $P_l \in \{X,Y,Z,I\}^{\otimes n}$, the first-order Suzuki trotterization is:
\begin{equation}
    e^{-i \sum_{l = 1}^{m} s_lP_l} = \prod_{l = 1}^{m} e^{-is_lP_l} + O(m^2s^2).
\end{equation}
where $s := \max_l \, s_l$. So for large $t$, to control error, we need to separate the evolution into multiple steps:
\begin{equation}
    e^{-i \sum_{l = 1}^{m} s_lP_l} = \left(\prod_{k = 1}^{m} e^{-i(s_l/r)P_l} \right)^r+ O(m^2s^2/r).
\end{equation}
where $r$ is the number of evolution steps.
There is also a second-order formula for smaller error
\begin{equation}
    e^{-i \sum_{l = 1}^{m} s_lP_l} = \left( \prod_{l = 1}^m e^{-i\frac{s_l}{2r}P_l} \prod_{l = m}^1 e^{-i\frac{s_l}{2r}P_l}  \right)^r + O(m^3s^3/r^2). \label{eq:suzuki-2ndr}
\end{equation}
So, if we set the trotterization error level at $\epsilon$, then 
we need to split the evolution into $r \in O\left( \frac{m^{3/2} s^{3/2}}{\sqrt{\epsilon}}\right)$ steps with the second-order formula.
By choosing $t = -\frac{s}{2r}$, Eq.\eqref{eq:exp_pauli_str} gives:
\begin{equation}
    e^{-i\frac{s}{2r}P} =  \cos\left(\frac{s}{2r}\right)I_n - i \sin\left(\frac{s}{2r}\right) P.
\end{equation}
As a result, Eq.\eqref{eq:suzuki-2ndr} becomes:
\begin{widetext}
\begin{equation}
    e^{-i \sum_{l = 1}^{m} s_lP_l} \approx \left( \prod_{l = 1}^m \left[ \cos\left(\frac{s_l}{2r}\right)I_n - i \sin\left(\frac{s_l}{2r}\right) P_l\right] \prod_{l = m}^1 \left[ \cos\left(\frac{s_l}{2r}\right)I_n - i \sin\left(\frac{s_l}{2r}\right) P_l\right]  \right)^r. \label{eq:exp_sukuzi}
\end{equation}
\end{widetext}
In this case, we approximate the exponential of the linear combination of the Pauli group matrices by another linear combination of Pauli group matrices, where the expectation of the latter one can be evaluated in a gate-based quantum computer easily. Note that in the $r = 1$ case, if we have $m$ terms in the power, we end up with at most $O(m^{2})$ terms after the transformation, which is exactly the scenario we had in H4 examples in the main text.

\bibliography{ref}
\end{document}